\documentclass[preprint,12pt]{aastex}

\usepackage{graphicx}
\usepackage{amsmath}
\usepackage{amssymb}

\newcommand{\EQ}[1]{Eq. \ref{EQ #1}}  
\newcommand{\FIG}[1]{Fig. \ref{FIG #1}}  
\newcommand{\qj}{\theta_{jet}}
\newcommand{\qobs}{\theta_{obs}}
\newcommand{\qb}{\theta_{beaming}}
\newcommand{\bj}{\beta_{jet}}
\newcommand{\vap}{v_{ap}}

\begin{document}

\title{The Apparent Size of GRB Afterglows as a Test of the
Fireball Model}
\author{Yonatan Oren, Ehud Nakar and Tsvi Piran}
\affil{
    The Racah Institute of Physics, The Hebrew University,
    Jerusalem, Israel, 91904}

\begin{abstract}
Taylor et al. (2004) reported recently on the first direct
measurement of the apparent size of a GRB afterglow. Here we
confront these observations with the predictions of the fireball
model. We use a simple model to study numerically the evolution of
the fireball and determine its apparent size, starting at the
early spherical expansion phase through the jet break and into the
Newtonian regime. We perform these calculations on the background
of a uniform interstellar medium and a massive stellar wind
environment. We find that the calculated apparent size at the time
of the measurements taken by Taylor et al. (2004) depends only
weakly on the chosen parameters (the jet opening angle, the
energy, and the external density profile). Thus it is reassuring
that the only possible outcome of the model, within the plausible
range of parameters, agrees very well with the data. These
measurements therefore present us with a strong test of possible
GRB models, which the fireball model passes successfully.
\end{abstract}

\section{Introduction}

Quite early in the study of the ``standard" afterglow fireball
model, it was realized that an observation of the size and
structure of the image of an afterglow would provide an almost
direct test of this model (Waxman 1997, Sari 1998, Granot Piran \&
Sari, 1999). Waxman (1997) calculated the ring shape of the
afterglow image while Sari (1998) calculated also the size of this
ring, noticing that the relativistic velocity of the afterglow
would result in a superluminal expansion. Until recently there
were no direct observations of the afterglow angular size.
Observations of decaying amplitude fluctuations in the radio
afterglow of GRB 970508 (Frail et. al., 1997) confirmed the
prediction of Goodman (1997) and enabled an estimate of the size
of this afterglow as $10^{17}$cm four weeks after the burst. This
observation agreed with an independent estimate based on the fact
that the system is optically thick to radio emission (Katz \&
Piran, 1997,Frail et. al., 1997). However, both methods are
somewhat model dependent and require assumptions on the source or
on the nature of the intergalactic medium. Still at that time a
direct measure of the size seemed as an unrealistic goal.

GRB 030329 is one of the most remarkable GRBs observed so far. It
was detected by HETE-II on March 29, 2003. Observations at many
different wavelengths followed, and a very bright afterglow
emission was detected over the whole spectrum. Its redshift,
$z=0.1685$ (Greiner et al. 2003),  places it as one of the nearest
GRBs whose redshift was firmly established. It is therefore
natural that it turned out to provide us with the first direct
measurement of the afterglow size and its expansion.

This difficult task was achieved by Taylor et al. (2004) who used
a large group of radio telescopes to resolve the radio afterglow
of GRB 030329 and demonstrated its relativistic expansion. They
find that the size of the afterglow is $0.07$mas (0.2 pc) 25 days
after the burst, and $0.17$mas (0.5 pc) 83 days after the burst.
The corresponding superluminal expansion is at a rate of 3-5c. At
a $15$GHz  observations taken 52 days after the burst Taylor et
al., (2004) detect an additional compact component at a distance
from the main component of $0.28 \pm 0.05$ mas (0.80 pc). As this
component was detected only once and then only in one band (it was
not detected at 22GHz at the same time or at  any other epoch at
any other band) we focus here on the observations of the main
component and we discuss only briefly the implications of this
additional component.

Before turning to the superluminal motion it is worthwhile to
mention several features of GRB 030329 that are relevant to this
discussion. During the first day the optical light curve of GRB
030329 looks rather typical, featuring an achromatic steepening
from $\alpha_1=0.87\pm 0.03$ to $\alpha_2=1.97\pm 0.12$ at
$T_j\approx 0.5\;$ days. Interpreting this
steepening as a jet-break, results in a half-opening angle of
$\theta_{j,0} \approx 0.07$rad, and a total $\gamma$-ray energy
output of $E_\gamma\sim 3\times 10^{49}\;$erg (Price et al. 2003).
After one day however, the optical light curve shows an unusual
variability, as several  re-brightening episodes increase the
observed flux by a large factor. Granot et al., (2003) interpret
these re-brightening events as refreshed shocks and estimate that
the total energy in these late shocks is about a factor of 10
larger than the initial energy seen in $\gamma$-rays, implying a
total energy of a few $10^{50}$ergs. Berger et al. (2003)  suggest
as an alternative explanation a two-component jet, where the wide
component has a half-opening angle of $\theta_{j,0} \approx
0.3$rad (corresponding to $T_j\approx 10\;$days) and an energy of
$\sim 3\cdot 10^{50}$ergs. Thus both models brings the total
energy of GRB 030329 to a few $10^{50}$ (the same level as most
GRBs) where the main difference in the context of this paper is
the break time.

Our goal here is to confront the fireball model with the new
observations of the size of the afterglow. This task is rather
simple for the spherical case, both at the early
ultra-relativistic phase (i.e. before the jet break; Sari, 1998),
and during the late transition to the non-relativistic phase. It
is  more complicated when we consider jets. Specifically, the rate
of the superluminal expansion depends on a combination of the
radius of emission, the Lorentz factor of the emitting region and
the direction of emission in the source frame. As we show later
the relation between the radius, the Lorentz factor and the angle
depends on the sideways expansion of the jet (past the jet break).
Thus, quite naturally, the superluminal behavior depends
critically on the full hydrodynamic behavior of the jet. In the
simplest model (Sari et al. 1999), the jet expands sideways
relativistically. On the other hand, numerical solutions (Granot
et al., 2001, Cannizzo etal., 2004) suggest that the expansion may
be much slower. A complete solution of the problem requires a
detailed hydrodynamic solution of the evolution of the jet
combined with a calculation of the time of arrival of the photons
from different emitting regions within the jet. Lacking such a
solution we present here simple analytic and numerical toy models
that enables us to explore the possible range of superluminal
expansion.

We show first that the angular size predicted by the fireball
model after a few weeks is rather robust (i.e does not depend
strongly on the details such as the external density profile and
the exact expansion velocity). Even so, detailed observations can
be used to distinguish between extreme expansion models  (i.e.
relativistic vs. no expansion). We use these results to
investigate GRB 030329.

In section 2 we present the theoretical framework for our
analysis, including the numerical model and the analytic
estimates. In section 3 we we discuss the results and compare them
with the observation of GRB 030329. Finally we discuss  the
implications of these predictions in view of of future
observations in section 4.

\section{Theory}

A point source moving toward a distant observer at an angle
$\theta$ to the line of sight would appear to move superluminally
across the plane of the sky (Couderc 1939). This has been noted as
an important feature of relativistically expanding radio sources
by Rees (1966). The apparent velocity $\vap$ is given in
\EQ{slm_beta} where $\beta$ and $\gamma$ are the source velocity
and Lorentz factor,
\begin{equation}\label{EQ slm_beta}
\frac{\vap}{c}=\frac{\sin(\theta)}{\beta^{-1}-\cos(\theta)}
\approx \frac{2\gamma}{\theta\gamma+{1}/{\theta \gamma}},
\end{equation}
which reduces in the relativistic limit and for $\theta$ close to
$1/\gamma$ to the simple self similar form given in the right
term.  The maximal value of $\vap \approx \gamma c$ is obtained
for $\theta \approx 1/\gamma$.

\begin{figure}[h]
\begin{center}
\includegraphics[width=10cm,height=6cm]{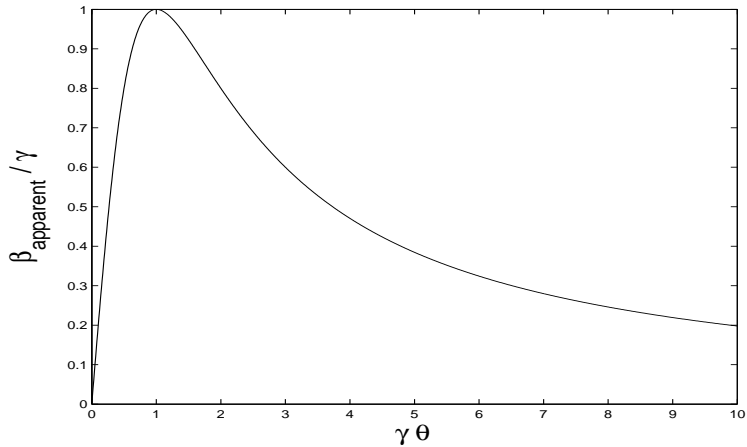}
\caption{The self similar apparent velocity of a relativistically
moving object. For a spherically expanding shell, the dominant
flux arrives from $\theta\approx \gamma^{-1}$, and is seen to have
the maximum superluminal velocity, $\vap=\gamma c$. \label{FIG
slm_beta}}
\end{center}
\end{figure}

For a spherically expanding fireball the dominant portion of flux
arrives from an angle of $1/\gamma$ relative to the line of sight,
making $1/\gamma$ a good approximation for the angular size of the
observed image.  This also happens to be the direction at which we
see the fastest superluminal motion (see \FIG{slm_beta}),
$\vap=\gamma c$. During this phase we can, in principle, track
directly the Lorentz factor of the shock through its image size.
Unfortunately,  the size of the afterglow is rather small at this
stage and this measurement is extremely difficult.

For a collimated and non-sideways expanding jet, we will observe
the $\theta=1/\gamma$ ring until $1/\gamma=\qj$, where $\qj$ is
the jet half opening angle, and then continue to observe emission
from the edge of the jet, $\theta=\qj$. This  will still appear
superluminal but with $\vap$ decreasing faster, reaching
asymptotically  $2\qj\gamma^2 c$. Thus, we expect to see a break
in the expansion velocity that is roughly simultaneous with the
break in the light curve. The break will be complete when the
$1/(\theta \gamma)$ term in the denominator will dominate over the
$\theta \gamma$ term, roughly over an order of magnitude in
$\gamma$.  The same qualitative behavior is seen for a
sideways-expanding jet but here a more cumbersome analysis is
required. As expected we find in this case an intermediate
decrease in $\vap$, between the two cases (spherical expansion and
a non sideways-expanding jet).

The superluminal velocity, $\vap$, cannot be measured directly.
Instead one measures the apparent size of the fireball:
\begin{equation}\label{EQ R}
 R_\bot=R \ \sin(\qobs) \ ,
\end{equation}
and its dependence on the observer time:
\begin{equation}\label{EQ T}
 T_{obs}=t-R \cos(\qobs) \ ,
\end{equation}
where $t$ is the time in the source frame. The angle
$\qobs(R,\Gamma,\qj)$ is the typical observing angle  (i.e. the
polar angle in the source frame from which the dominant amount of
flux reaches the observer). It satisfies $\qobs={\rm
min}(\qj,\qb$), where $\qj$ is the jet opening angle and  $\qb$ is
the relativistic beaming angle,
$\qb=\tan^{-1}[(\gamma^2-1)^{-1/2}]$.

In principle one needs two dimensional hydrodynamic calculations
in order to follow the evolution of the jet and to obtain $R(t)$,
$\gamma(t)$ and $\qj(t)$ as functions of $t$. Lacking such
computations we construct a simple hydrodynamic model for the
expanding jet. Within this simple model, which follows the common
assumptions made in analyzing jets we approximate the full two
dimensional evolution by a one dimensional solution in the radial
direction. Instead of calculating the exact lateral expansion we
assume a sideways propagation speed $\beta_{jet}$. We examine the
upper and lower limits  $0\le \beta_{jet}\le 1$as well as  two
physical models for $\beta_{jet}$. We also assume that the jet is
uniform in its lateral direction and that its shape remains
spherical.

Ignoring the kinetic energy involved in the lateral expansion we use the
adiabatic equation for the energy transferred to the ISM by the
blast wave:
\begin{equation}\label{EQ adiabatic}
E=C(\gamma)(\gamma^2-1) Mc^2,
\end{equation}
where $\gamma$ is the (radial) Lorentz factor of the freshly
shocked material ($\sqrt{2}\gamma=\Gamma$, the Lorentz factor of
the shock front) and $M$ is the accumulated mass swept by the
blast wave. $C(\gamma)$ is a numerical factor of order unity. It
is a constant in the relativistic regime. Its $\gamma$ dependence
enables us to extend the evolution of the fireball into the
Newtonian regime. We find its value using the shock crossing
conditions. Adding to \EQ{adiabatic} a model for the external
density, $\rho(R)\propto R^{-k}$, ($k=0$ for ISM and $k=2$ for
wind), and a model for the sideways expansion of the jet which we
will discuss later, we can calculate the hydrodynamic evolution of
the jet.

Within our simplified model the jet evolution is described by a
set of four ordinary differential equations:
\begin{equation}\label{EQ jet_ode}
\begin{array}{l}
 \ \ \ \ \ \frac{dR}{dt}=c\sqrt{1-\Gamma^{-2}} \\ \\
 \ \ \ \ \ \frac{d\qj}{dt}=\frac{\bj}{R \Gamma} \\ \\
 \ \ \ \ \ \frac{dM}{dR}=4\pi \tilde \rho R^{2-k} [1-\cos(\qj)]  \\ \\
 \ \ \ \ \
\frac{d\gamma}{dM}=-[\frac{2\gamma}{\gamma^2-1}M+\frac{M}{C}
\frac{dC}{d\gamma}]^{-1},
\end{array}
\end{equation} \\
where $c=1$ and
\begin{equation}
\tilde \rho =
\left\{%
\begin{array}{ll}
    n_0 m_p, & \hbox{for ISM;} \\
     5\cdot10^{11} gr/cm A_*, & \hbox{for a wind with k=2.} \\
\end{array}%
\right.
\end{equation}
We close this system with expressions for $\Gamma(\gamma)$ and
$C(\gamma)$ derived from the (radial) shock crossing conditions,
and a model for $\bj$, the sideways expansion velocity of the jet
in the shock frame. We consider four cases: (i) $\bj=0$,
representing models in which the jet does not expand or expands
very slowly. There are some indications from numerical works
(Granot et al., Cannizzo et al., 2004, Kumar \& Granot 2003) for
such a behavior. (ii) $\bj=1$, which represents the limit of
ultra-relativistic expansion. This approximation was used, for
example, by Sari, Piran, \& Halpern(1999) for their calculations
of the post jet break light curves.  These two cases serve as
lower and upper limits on the expansion rate.  The two other, more
physical models, feature a variable expansion speed that follows
the physical conditions in the shocked matter: (iii)
$\bj=\beta_s$, the sound speed (see for example Panaitescu and
Meszaros, 1999).  $\beta_s$ approaches $1/\sqrt{3}$ in the limit
of relativistic shocks. It vanishes in the limit of a cold,
Newtonian shock. (iv) $\bj=\beta_T$, the thermal velocity of the
shocked material. If we consider the lateral dimension alone, the
low density and pressure outside the jet will give rise to a
rarefication wave at the edges that will propagate with
approximately $\beta_T$. At the relativistic limit $\beta_T$
approaches 1, and it vanishes in the Newtonian limit.

The evolution from the observer point of view proceeds in three
asymptotic stages: (i) The pre-break (spherical) stage, $\gamma
\gg \qj^{-1}$, where the dynamics are similar to spherical
expansion and the visible part of the jet is
$\theta_{obs}=1/\gamma$. (ii) The relativistic post-break stage,
$\qj^{-1}\gg\gamma \gg1$, when  the visible region is determined
by the edge of the jet $\theta_{obs}=\qj$ (this is the only stage
that is sensitive to $\bj$) (iii) The Newtonian stage when $\gamma
\sim 1$, where the expansion is Newtonian and the entire fireball
is visible.

It is possible to derive analytical expressions for the behavior
of the observed size $R_\bot$ in these three asymptotic stages of
the afterglow. We do this by employing the adiabatic equation and
different approximations for the value of $\theta_{obs}(T)$. Our
results agree with those of  Sari (1998) for spherical propagation
into an ISM and with those of Galama et al. (2003) for a jet
expanding relativistically into an ISM ($\bj=1$). We are also in
agreement with the results of Galama et al. (2003) for the
spherical propagation into a wind at asymptotically late times
(before the newtonian phase). In the first stage we take the
spherical adiabatic equation and $\theta_{obs} = 1/\gamma$. In the
second stage, if $\beta_{jet}$ vanishes than the spherical
adiabatic equation still holds but $\theta_{obs}(T) = \theta_0$.
If $\bj \sim 1$ the adiabatic equation implies $R \sim const$ and
$\theta_{obs} \sim 1/\gamma$. In the last, Newtonian, stage the
emission is not beamed anymore and the evolution becomes simply
the Sedov-Taylor solution (Sedov 1946, Taylor 1950, Von Neumann
1947). At this stage the the apparent expansion speed naturally
becomes sub-luminal. We summarize the different approximations in
table 1 (Note that here and elsewhere in the paper $E$ stands for
the total energy of the flow and $E_{iso}$ for the isotropic
equivalent energy.):

\begin{table}[h]
\begin{center}
\begin{tabular}{||c|c|c|c||}
\hline \hline
\multicolumn{2}{||c|}{} & ISM & WIND \\ \hline
Pre Break & by  $E_{tot}$ &
$5\cdot 10^{16}(\frac{E_{51}}{n})^{1/6} T_{j,day}^{-1/8}
T_{day}^{5/8}$
&
$4\cdot 10^{16}(\frac{E_{51}}{A_*})^{1/2} T_{j,day}^{-1/4}
T_{day}^{3/4}$ \\ \cline{2-4}
(Spherical) & by $E_{iso}$ &
$3.5\cdot 10^{16}(\frac{E_{52,iso}}{n})^{1/8} T_{day}^{5/8}$
&
$10^{16}(\frac{E_{52,iso}}{A_*})^{1/4} T_{day}^{3/4}$ \\
\hline \hline
Post Break & $\beta_j \sim 1$ &
$6\cdot 10^{16}(\frac{E_{51}}{n})^{1/6} T_{day}^{1/2}$
& - \\ \cline{2-4}
&  $\beta_j << 1$ &
$8\cdot 10^{16}(\frac{E_{51}}{n})^{1/6}  T_{j,day}^{1/4} T_{day}^{1/4}$ &
$7\cdot 10^{16}(\frac{E_{51}}{n})^{1/6}  T_{day}^{1/2}$ \\
\hline \hline
Newtonian & &
$10^{18}(\frac{E_{51}}{n})^{1/5} T_{year}^{2/5}$
&
$10^{18}(\frac{E_{51}}{A_*})^{1/3} T_{year}^{2/3}$ \\
\hline \hline
\end{tabular}
\end{center}
\caption{Analytic approximations in the 3 asymptotic stages for the
  observed size $R_\bot$. The absence of an entry for the post break
  stage in wind is due to the fact that the break in this case is so
  gradual that the analytic solution is not a good approximation at
  any point between the jet break and the Newtonian regime.}
\end{table}

The transition between the asymptotic stages must be calculated
numerically using \EQ{jet_ode}. The transition from the first to
the second stage (i.e. the jet break of $R_\bot$) has similar
properties to the jet-break in the light curve. It is a result of
two effects, the sideways expansion (that is relevant only when
$\bj$ is relativistic) and the decrease in the relativistic
beaming that brings (when $1/\gamma> \qj$)  the edge of the jet
into view. The sideways expansion becomes important when
$1/\gamma\theta_0=\qj(T=0)$. This takes place at the canonical
estimate of the break time (Sari et. al. 1999, Chevalier \& Li
2000)
\begin{equation}
T_{jet} \approx \left\{
\begin{array}{l}
 \ 1 (E_{51}/n)^{1/3} (\theta_0/0.1)^2 (1+z) \ days\ \ \ ISM\\ \\
 \ 1 (E_{51}/A_*) (\theta_0/0.1)^2 (1+z) \ days \ \ \ WIND\\ \\
\end{array}
\right.
\label{tjet}
\end{equation}
 The edges of the jet are actually observed only when
$1/\gamma(T)=\qj(T)$. When the jet expansion is negligible $\qj(T)
\approx \theta_0$ at any $T$ and both effects occur the the same
time. When $\bj$ is relativistic then the edges of the jet are
observed later than $T_j$, in an ISM by a factor of few and in a
wind by more then an order of magnitude. Kumar and Panaitescu
(2000) explore the effect of these two phenomena on the break in
the light curve assuming a relativistic $\bj$. They find that the
beaming effect is more dominant, and that the break takes about
one decade in time in an ISM and more than two decades in time in
a wind. Similar behavior is observed in the break of $R_\bot (T)$.
As a result of the very long break in a wind environment (when
$\bj \sim 1$), the second stage never reaches its asymptotic
behavior and it is impossible to find an accurate analytic
approximation to this stage.  Finally, We find that when $\bj \ll
1$ the break take place at $T_j$ and it takes about one decade in
time both for ISM and wind. The transition to Newtonian stage
takes about one time decade in an ISM and two decades in a wind.

The break is also affected by the viewing angle of the jet. The
observation of a prompt emission of $\gamma$ rays implies that our
viewing angle is smaller than $\theta_0$, but we are still
uncertain of the angle between the axis of the jet and the line of
sight. If we are not viewing the jet exactly on axis, the
$1/\gamma$ ring will cross the edge of the jet at different times
for different azimuthal angles, an effect which will alter the
shape of the break. We ignore this effect here for the sake of
simplicity, as it should not affect our results in a significant
manner.

\section{Application to GRB 030329}

The key measurements made by Taylor et al. (2004) on GRB 030329
indicate that the angular size of the emitting region in the radio
band is $0.04-0.11$mas $25$ days after the burst, and between
$0.13-0.21$mas $83$ days after the burst, with nominal values
$0.077$ and $0.172$ mas at these times. Another measurement at $51$
days after the burst gives a $2\sigma$ upper limit of $0.1$mas
(see, e.g., \FIG{ISM_bs}). The angular distance to the burst
corrected for cosmological effects is estimated as $589 Mpc$, which leads
to a radial size of $6.7\cdot 10^{17} cm, 8.7\cdot 10^{17} cm$ and
$15\cdot 10^{17} cm$ respectively for the three measurements.
Given a power law expansion of the observed size with time,
$R_\bot \propto T^\alpha$, these measurements constrain the
exponent to be $0.5 \lesssim \alpha \lesssim 1.1$ in the
neighborhood of the observations. Taylor et al. (2004) fit the
observations using the analytical asymptotic equations of
$R_\bot(T)$ given by Galama et al. (2003) for the spherical stage
and the post-break stage (with $\bj \approx 1$). They find that
all the models can fit the observations. They find the best fit
values of $E_{51}/n$ in ISM (or $E_{51}/A_*$ in a wind) to be
$\sim 10$ for a spherical wind and a jet with $T_j=10$days, and
$\sim 2$ for a jet with $T_j=0.5$days.

\begin{figure}[h]
\begin{center}
\includegraphics[width=10cm,height=6cm]{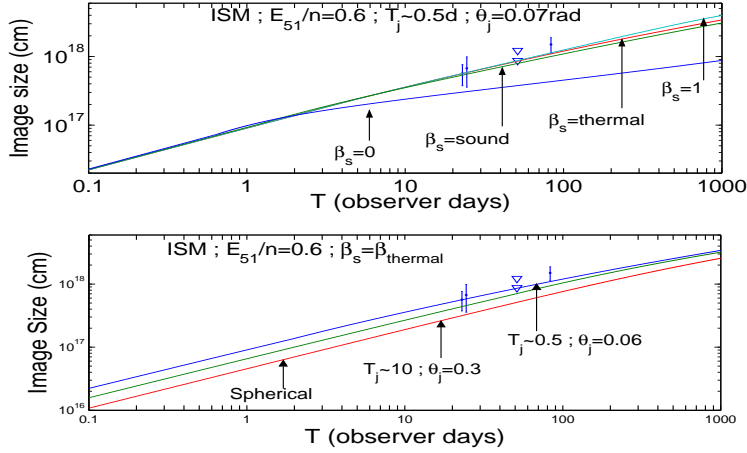}
\caption{{\it upper pannel}: $R_\bot$ as a function of $T_d$ for
different sideways expansion models in ISM. The energy to external
density ratio $E/n=0.6\cdot 10^{51}erg \ cm^3$ and $\theta_0=0.06
rad$. {\it Lower pannel}: $R_\bot$ as a function of $T_d$ for
different opening angles in ISM, with constant $E/n=0.6\cdot
10^{51}erg \ cm^3$. $T_j$ is in days, $\beta_s=\beta_{thermal}$ }
\label{FIG ISM_bs}
\end{center}
\end{figure}

\begin{figure}[h]
\begin{center}
\includegraphics[width=10cm,height=6cm]{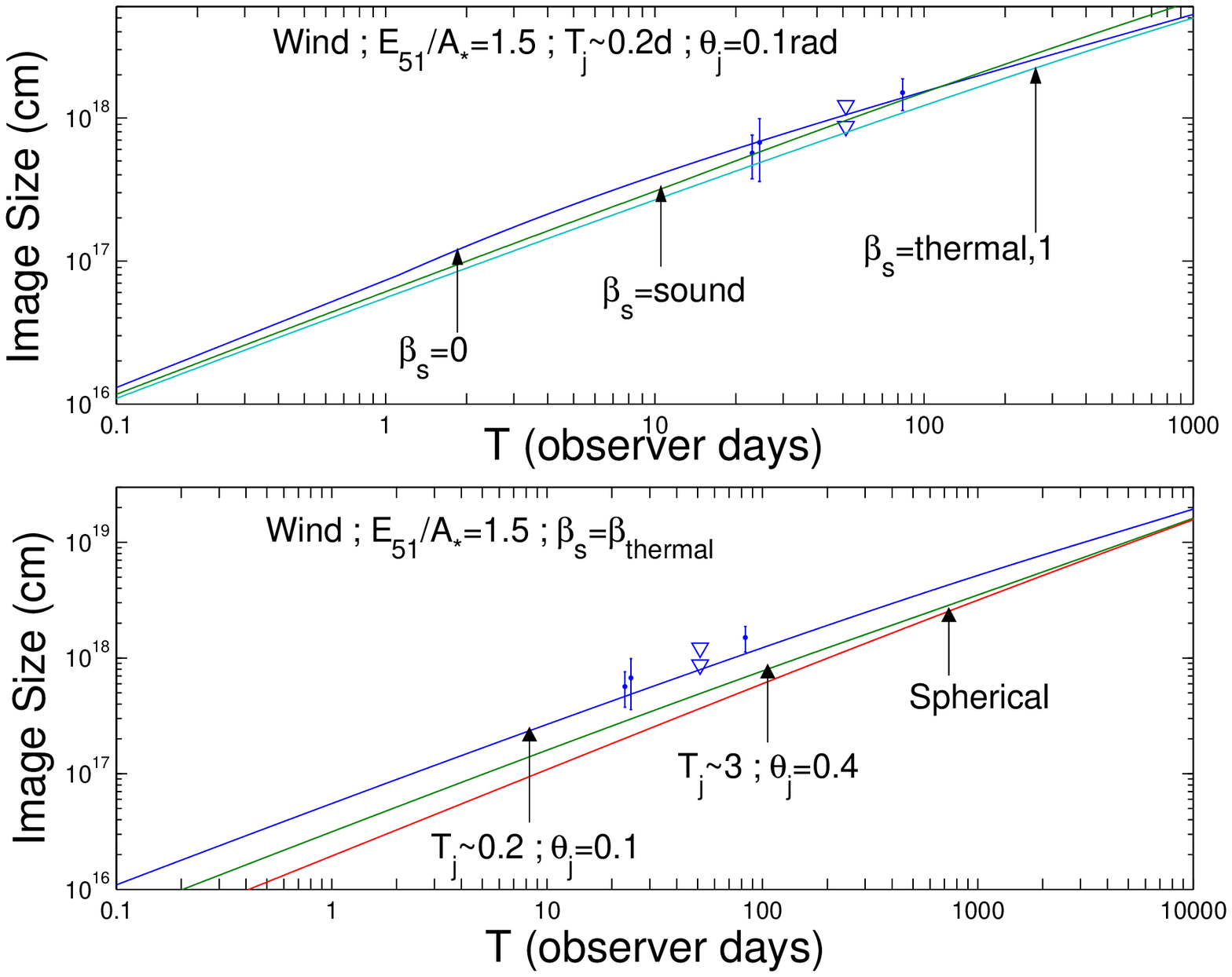}
\caption{{\it upper pannel}: $R_\bot$ as a function of $T_d$ for
different sideways expansion models in a wind.$E/n=0.6\cdot
10^{51}erg \ cm^3$ and $\theta_0=0.06 rad$. {\it Lower pannel}:
$R_\bot$ as a function of $T_d$ for different opening angles in a
wind, with $E/n=0.6\cdot 10^{51}erg \ cm^3$. $T_j$ is in days,
$\beta_S=\beta_{thermal}$.} \label{FIG WIND_bs}
\end{center}
\end{figure}

We used our numerical model to try and confront these measurements
with the predictions of the fireball model. The first and the most
remarkable result is that the angular size ($\sim 10^{18}$ cm) and
its increment index $0.5 \lesssim \alpha \lesssim 1.1$ are similar
to the predictions for the most typical GRB parameters
($E51/n=0.1-10$) with little dependence on the beaming factor,
expansion velocity and external density profile, with the
exception of a non expanding jet in ISM. The rather constant
values of $\alpha$ for the different models ($\alpha=0.5-0.75$
when a non expanding jet in ISM is excluded), and the weak
dependance of $R_\bot$ on $E/n$ leave a very narrow range of
possible results, making the measurement of $R_\bot(T_{obs})$ a
very strong test of the fireball model.

However, the same feature that make this measurement such a strong
consistency check on the fireball model also makes it a rather
poor instrument for distinguishing between the different models.
It is evident that the current measurements do not allow us to
rule out any of the suggested models (\FIG{ISM_bs},
\FIG{WIND_bs}), except for the non expanding jet in ISM where
$\alpha \approx 0.25$. This value of $\alpha$ makes any such
configuration with $T_{jet}<25$days inconsistent with the data.
This result is the first observational hint of the value of $\bj$.
On the other hand even a model with $\bj=\beta_s$ gives almost the
same results as $\bj=1$.

The best fit for the observations in an ISM where $\bj$ is
relativistic is obtained with the following values of $E_{51}/n$:
$\sim 0.5(T_j \sim 0.5$days), $\sim 1(T_j \sim 10$days) and $\sim
10$ if the ejecta is spherical. In a wind environment all the
different models require $E_{51}/n \sim (1-a\; few)$ to fit the
observations. If we interpret, however, the sharp break in the
optical afterglow at $0.5$days to be a jet break, than we can
argue against the wind scenario with $\bj \sim 1$, as it would
imply a much more gradual break spanning roughly two orders of
magnitude in time (Kumar \& Panaitescu 2000). Another interesting
feature of the results is that the estimates of the burst energy
out of the optical (Granot et al. 2003) and the radio (Berger et
al. 2003) observations, a few $\times 10^{50}$ergs with the
observed break time ($0.5$day or $10$days) is also the energy
value that provides the best fit to $R_\bot(T_{obs})$ (assuming
the canonical value of $n=1 \rm cm^{-3}$). Even though the weak
dependance of $R_\bot$ on $E$ (a power of $1/6$) makes it a blunt
tool to probe the energy, this result is reassuring.

Note that the best fit values of Taylor et. al. (2004) of $E/n$
for the beamed models ($T_j$ = 0.5 or 10days) are an order of
magnitude larger than our best fit values. The reason is the sharp
break taken in the modelling of Galama et. al. (2003) compared to
the gradual break in the numerical results. This deficiency
however is within the uncertainty of the model due to the weak
dependance of $R_\bot$ on $E$ (i.e. a difference of a factor ten
in the energy reflects only a difference of a factor 1.4 in
$E/n$).

Finally we note that Taylor et al. (2004) report a single
detection of a secondary component at 15GHz after $51$ days at a
distance of $\sim 0.8$pc, implying  an average apparent velocity
of $19c$ (apart for the detection of the main afterglow
component). It is important to note this component was not
detected at any other epoch, and more importantly it was not
detected in a less sensitive observation at 22GHz taken at the
same time (while the detection at 15GHz was at a level of
$>20\sigma$). Thus, we think that it is premature to contrive a
theory based  on this observation.

Still, we address this observation at face value, assuming that the
source of the emission was ejected at the time of the burst from
the same location of the main component. The apparent velocity of
a source moving at the speed of light at an angle $\theta$ to the
observer is $\vap = 2/\theta$. Therefore, the source of the
secondary component must be at an angle $\le 0.1rad$ implying
$R=R_\bot / \sin (\theta)\ge  2\cdot 10^{19}cm$ and $\Gamma \ge
10$ at this radius (see Eq. \ref{EQ slm_beta}) . This combination
is incompatible with the fireball model with any reasonable
parameters. Thus, if the exitance of such a component would be
confirmed in the future it will probably require a theoretical
revolution. However, as we stressed earlier, it is premature to
reach such a drastic conclusion, particularly in view of the
remarkable success of this theory in explaining the main
component.

\section{Conclusions}

Until recently no direct measurement of the size of the emitting
region in a GRB afterglow was available, due to the large
distances and relatively small sizes involved. In a situation
where most of the information we have on the sources of GRB's is
indirect and often ambiguous, such a measurement is extremely
important as it allows us to test directly some of the predictions
of existing models. This goal has been achieved with the afterglow
of GRB 030329 (Taylor et al. 2004), and in this work we tried to
test the compatibility of the fireball model with this new data
and to derive some general predictions for future observations.

Relativistically expanding fireballs cast an image on the plane of
the sky that grows superluminally with the observer time. A
detailed modelling of this scenario would require a full 3D, or at
least 2D relativistic hydrodynamic code. Lacking such a code we
made a simpler analysis that captures the essential features of
interest. We consider a jet described by a radial shock, bounded
to a cone of opening angle $\qj$, which grows at a rate determined
by the properties of the ejecta in its rest frame. This simple
scheme enables us to model the superluminal expansion of GRB
afterglows in general and the afterglow of GRB 030329 in
particular and try to constrain some of its parameters.

Our main result is that both the value and the time evolution of
$R_\bot$ depends rather weakly on the exact variant of the
fireball model and its parameters. We find that the main afterglow
component of GRB 030329 fits remarkably well the predictions of
the fireball model. Specifically we find that the exact values of
the current estimates in the literature on the energy and opening
angle of GRB 030329 give excellent fits to the data. There is only
one variant of the fireball model which appears to be rejected by
the observations, the non expanding jet in ISM. We consider the
fact that a direct measurement confirms a robust prediction of the
fireball model as one of the model's most important achievements.

We conclude with the reflection that while the direct observations
of angular size are maybe a strong test of the fireball model,
they cannot be used (at least in their current quality) as a tool
to distinguish between different scenarios. However, if in the
future we will have more accurate and more frequent observations
then it might be possible to use them in order to achieve this
task as well.

The research was supported in part by a US-Israel Binational
Science Foundation.

\end{document}